\documentclass[9pt,twocolumn,twoside]{osajnl}
\usepackage{booktabs}
\usepackage{threeparttable}
\usepackage{stfloats}
\usepackage{multirow}

\journal{jocn} 

\setboolean{shortarticle}{false}

\title{Real-Time Burst-Mode Digital Signal Processing for Passive Optical Networks}

\author[1,*]{Ji Zhou}
\author[1]{Kainan Wu}
\author[2]{Haide Wang}
\author[1]{Jinyang Yang}
\author[1]{Weiping Liu}
\author[3]{Junwen Zhang}
\author[4]{Changyuan Yu}
\author[5]{Xiangjun Xin}
\author[5,6]{Liangchuan Li}

\affil[1]{Department of Electronic Engineering, Jinan University, Guangzhou 510632, China}
\affil[2]{School of Cyber Security, Guangdong Polytechnic Normal University, Guangzhou 510665, China}
\affil[3]{Department of Communication Science and Engineering, Fudan University, Shanghai 200433, China}
\affil[4]{Department of Electrical and Electronic Engineering, The Hong Kong Polytechnic University, Hong Kong SAR, China}
\affil[5]{Laboratory of Broadband Optical Network Algorithm and ASIC, Beijing Institute of Technology (Zhuhai), Zhuhai 519088, China}
\affil[6]{liliangchuan@bitzh.edu.cn}
\affil[*]{Corresponding Author: zhouji@jnu.edu.cn}

\begin{abstract}
Driven by the ever-increasing capacity demands, the 50G passive optical network (PON) is maturing gradually. One of the main challenges for the 50G PON is implementing burst-mode digital signal processing (BM-DSP) for the burst upstream signal. In this paper, we demonstrate a real-time BM-DSP for burst reception of 25Gbit/s on-off keying signal to meet the asymmetric-mode 50G PON demand. The real-time BM-DSP includes the BM frequency-domain timing recovery and BM frequency-domain equalizer, which can be fast converged based on the 42ns designed preamble. Meanwhile, the simplified implementations for fast-Fourier-transform, minimum-mean-square-error, and decision-directed least-mean-square-error algorithms decrease the DSP resources by 28.57$\%$, enabling the loading of real-time BM-DSP in the field programmable gate array with the limited DSP resources. The real-time implementation of BM-DSP can guide the design of application-specific integrated circuits for 50G PON.
\end{abstract}

\setboolean{displaycopyright}{false} 

\begin{document}

\maketitle

\section{Introduction}
The passive optical network (PON) development has been driven by artificial intelligence, mobile internet, and high-definition video \cite{rizzelli2024planning, luo2024beyond, 10403903, xing2024low, zhang2022coherent}. The International Telecommunication Union Telecommunication Standardization Sector (ITU-T) has released the standards for high-speed PON with line rates of 50Gbit/s (50G PON) to meet the ever-increasing capacity demands \cite{zhang2020progress, bonk202250g, zhang2021carrier}. To recover the downstream signal with line rates of 50Gbit/s and upstream signal with line rates of 12.5Gbit/s, 25Gbit/s, and 50Gbit/s, the 50G PON first adopts digital signal processing (DSP) to compensate for inter-symbol interference caused by the limited bandwidth and chromatic dispersion \cite{li2020dsp, liang2022dsp, torres2022overview}. The downstream broadcast signals received by all the optical network units (ONUs) are continuous, which can be processed by the continuous-mode DSP. Since the beginning of the commercial PON, statistical-multiplexing time-division multiple access (TDMA) ensures capacity and quantity for the subscribers \cite{kani2020current, dhaini2013energy, skubic2009comparison}. Therefore, the upstream signals received by the OLT are burst from each ONU. Burst-mode DSP (BM-DSP) is required to process the upstream burst signal in the 50G PON \cite{zhou2024burst, zhang2022demonstration, yin2021linear}.

In the continuous-mode DSP, the timing recovery and channel equalizer with feedback updating algorithms have long convergence times, which are the main obstacles for processing the upstream burst signal \cite{wang2023fast, van2019digital, teixeira2020dsp}. Therefore, many efforts have been committed to the BM timing recovery (TR) and BM channel equalizer for BM-DSP. The BM-DSP usually employs feed-forward algorithms based on a specially designed preamble to estimate the sampling phase offset (SPO) for TR and tap coefficients for the channel equalizer \cite{koma2021fast, matalla2022real, matalla2024comparison}. After the feed-forward algorithms, it will be switched to the feedback algorithms for updating the SPO and tap coefficients. Therefore, there are the feed-forward initialized algorithms, feedback updating algorithms, and many conditional judgments for algorithm switching, which require real-time implementations to verify the feasibility of BM-DSP \cite{7778247, li2013100, 10153945, xing2023first, matalla2021hardware}. Unfortunately, most of the existing research lacks the implementation details for the real-time BM-DSP. 

In this paper, we implement the real-time BM-DSP based on the designed preamble including BM frequency-domain TR (BM-FDTR) and BM frequency-domain equalizer (BM-FDE) for the 25Gbit/s upstream burst on-off keying (OOK) signal, meeting the requirement of the asymmetric-mode 50G PON. Meanwhile, we give detailed information about the flows, parallelism degrees, and delays for the real-time BM-DSP. The main contributions of this paper are as follows: 
\begin{itemize}
\item We implement real-time BM-DSP for 25Gbit/s OOK burst reception of asymmetric-mode 50G PON, mainly including the BM-FDTR and BM-FDE with a fast convergence based on the 42ns designed preamble.
\item The improved real-time implementations for fast Fourier transform (FFT), minimum-mean square error (MMSE), and decision-directed least-mean-square (DD-LMS) algorithm decrease the DSP resources by 28.57$\%$, enabling the loading of real-time BM-DSP in the field-programmable gate array (FPGA) with the limited DSP resources.
\end{itemize}

The remainder of this paper is organized as follows. The designed preamble and implementation for real-time BM-DSP are shown in Section \ref{Section2}. In Section \ref{Section3}, the experimental setup and results for the real-time 25Gbit/s OOK are demonstrated. The paper is concluded in Section \ref{Section4}. 

\section{Implementation of Real-Time BM-DSP} \label{Section2}
This section demonstrates the overall frame structure and BM-DSP flow. Then, detailed information about the main algorithm modules' flows, parallelism degrees, and delays is given.

\subsection{Frame Structure and BM-DSP Flow}
\begin{figure}[!t]
\centering
    {\includegraphics[width = \linewidth]{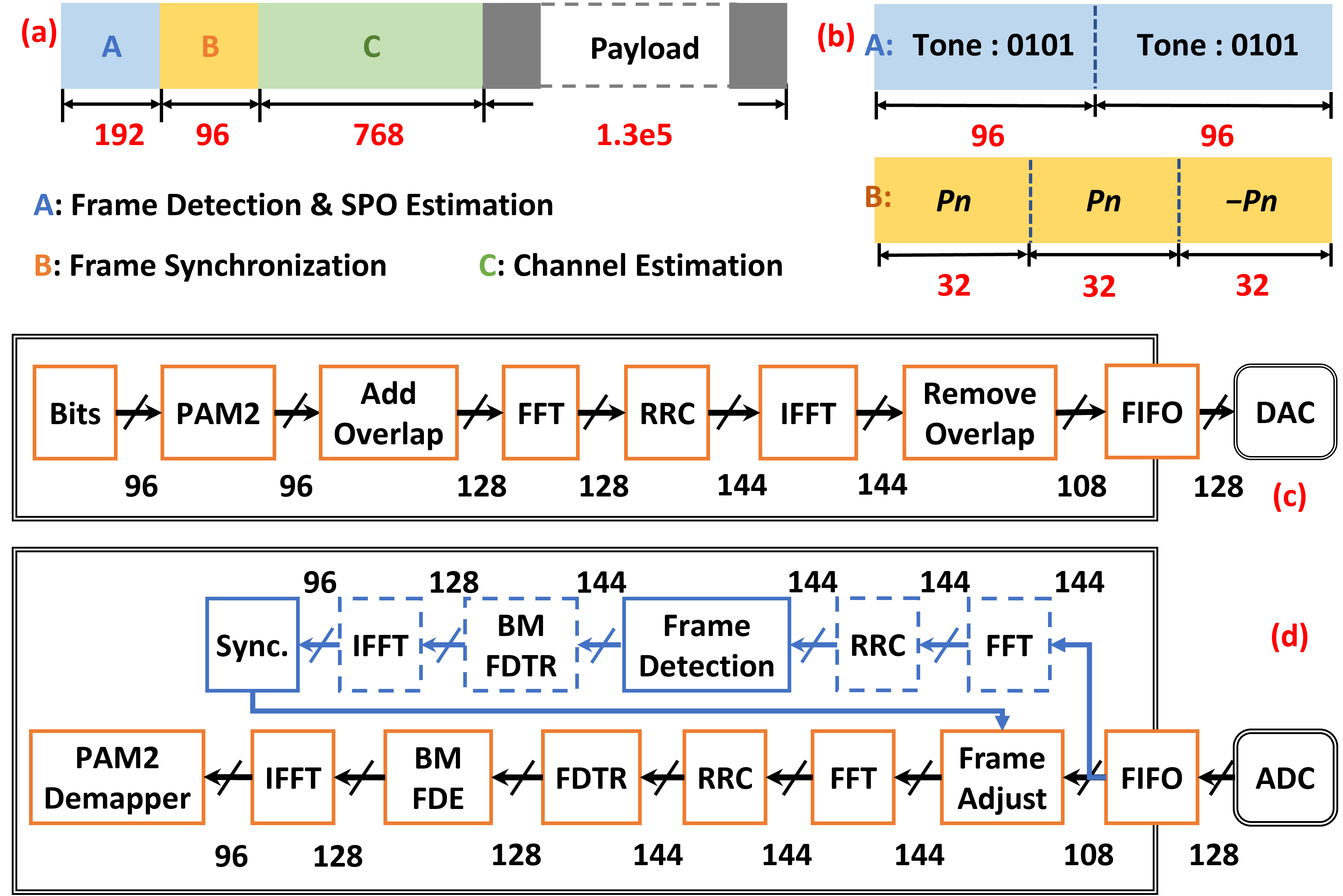}}
\caption{(a) The frame structure for burst reception of 50G PON. (b) Detailed design of Preamble A and Preamble B. (c) The flow of the real-time signal generation at the transmitter side. (d) The flow of the real-time BM-DSP at the receiver side.}
\label{fig:dspflow}
\end{figure}

Figure \ref{fig:dspflow} (a) shows the frame structure with the designed preambles to realize the fast convergence for the real-time BM-DSP. The designed preamble has three parts: A, B, and C. Preamble A with 192 symbols is designed for frame detection and SPO estimation. Preamble B with 96 symbols is designed for frame synchronization. Preamble C with 768 random symbols is used to estimate the initial tap coefficients of the BM-FDE. Therefore, the total length of the designed preamble is 1056 (i.e. $\sim$ 42ns @ 25Gbit/s OOK signal), and the payload length is set to $1.3\times 10^5$. Figure \ref{fig:dspflow} (b) depicts the design for Preamble A and Preamble B. Preamble A consists of the repeated sequence $[0,1]$ in the time domain, which presents as two tones at half of the baud rate in the frequency domain (i.e. $R_\text{s}/2$ where $R_\text{s}$ denotes the baud rate). Preamble A occupies two full clock cycles, where the first beat is used for frame detection, and the second one is used for SPO estimation. Preamble B comprises three sets of identical 32-symbol random sequences $Pn$ multiplied by coefficients of $[1,1,-1]$, which can ensure enough power of synchronization peak to achieve the accurate frame synchronization.

Figure \ref{fig:dspflow} (c) shows the flow of the real-time signal generation at the transmitter (Tx) side. Firstly, bits are mapped into two-level pulse amplitude modulation (PAM2) symbols in parallel. The parallelism degree is set to 96. After adding a 32-symbol overlap, the time-domain signal is transformed into the frequency-domain signal by a 128-point FFT. By combining the first 72 frequency points and the last 72 frequency points into 144 frequency points, the signal at 1 sample per symbol (sps) is resampled to 1.125 sps. The 144 frequency points are pulse-shaped by a root-raised cosine (RRC) filter with a roll-off factor of 0.1. After 144-point inverse FFT (IFFT), the parallelism degree of the time-domain signal is converted to 108 by removing the 36-overlap (i.e. $32\times 1.125$)  symbols. Finally, after the first-in-first-out (FIFO) for parallelism alignment, the digital signal with a parallelism degree of 128 is converted to the analog signal by a digital-to-analog converter (DAC).

Figure \ref{fig:dspflow} (d) depicts the flow of the real-time signal generation at the receiver (Rx) side. First of all, the analog signal is converted to the digital signal with a parallelism degree of 128 by an analog-to-digital converter (ADC). After the FIFO for parallelism alignment, the parallelism degree is changed to 108. A 36-symbol overlap is added to obtain one signal beat with a parallelism degree of 144. After 144-point FFT, frame detection based on Preamble A is performed to confirm the arrival of the burst signal. Then, the BM-FDTR based on Preamble A can quickly estimate and compensate for the SPO. The parallelism degree is changed to 128 by removing the roll-off frequency points. After 128-point IFFT and removing the 32-symbol overlap, the frame position is extracted by the frame synchronization based on Preamble B at 1 sps. After frame adjustment using the frame position, a 36-symbol overlap is added to make up one synchronous signal beat with a parallelism degree of 144. After the FDTR, the synchronous signal beats are fed into the BM-FDE. The BM-FDE based on Preamble C can quickly estimate the coefficients and compensate for the channel distortions. After 128-point IFFT and removing the 32-symbol overlap, the PAM2 symbols are demapped to recover the bits.

\subsection{Real-Time Frame Detection and BM-FDTR}
\begin{figure}[!t]
\centering
    {\includegraphics[width = \linewidth]{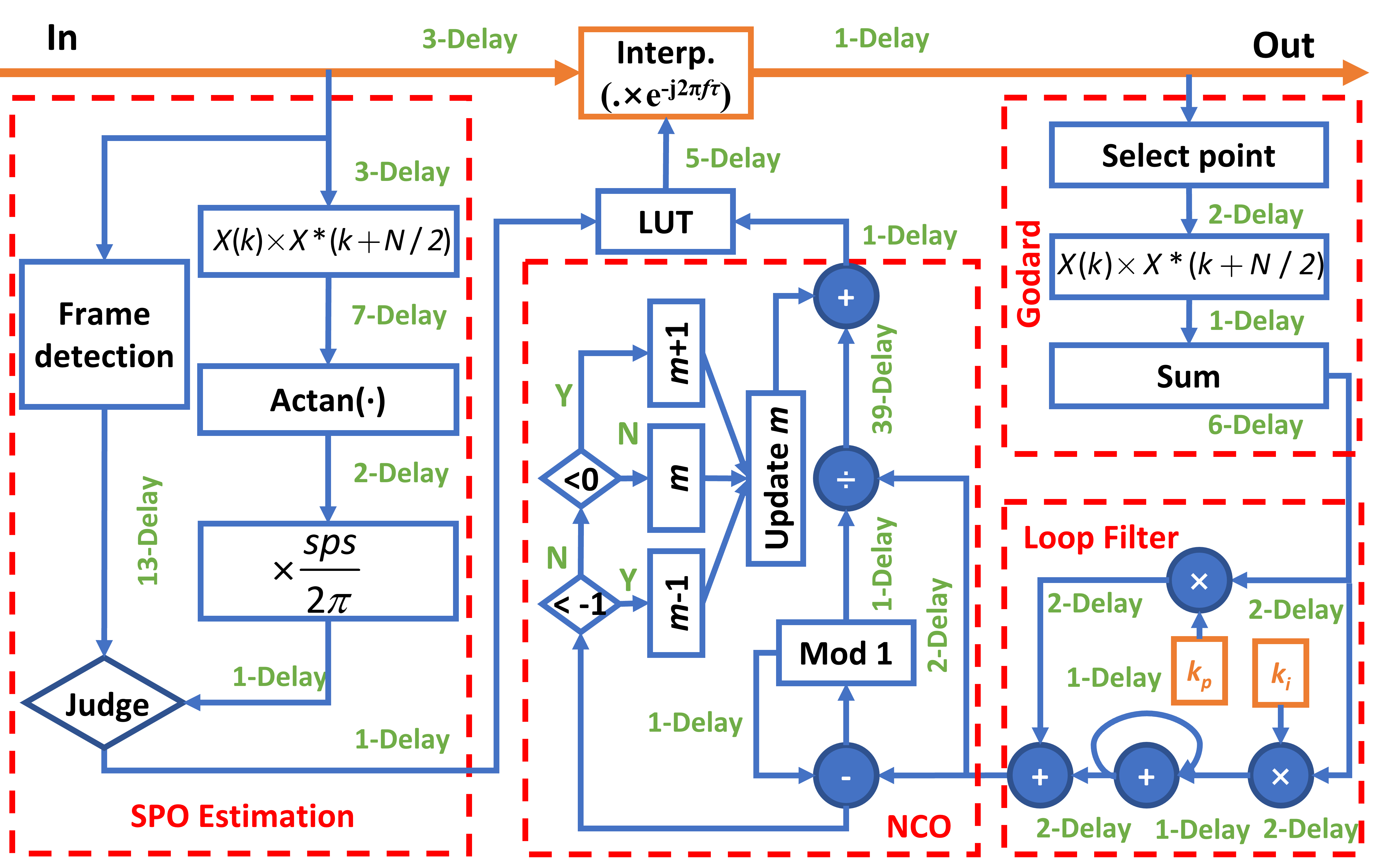}}  
\caption{The real-time implementation of frame detection and BM-FDTR with SPO estimation based on Preamble A.}
\label{fig:tr}
\end{figure}

Figure \ref{fig:tr} shows the real-time implementation of frame detection and BM-FDTR with sampling phase offset estimation based on Preamble A. The frame detection is performed to find the power peak of the Preamble A, which can confirm the arrival of the burst signal. If the preset frequency points have the maximum power, it can be considered that Preamble A is detected and the burst frame arrives. This paper uses the frequency points around $N/(2\times sps)$ and $N-N/(2\times sps)$ where $N$ is the FFT size. A comparison based on a tree structure search to find the frequency points with the maximum power. The tree structure search has 7 layers, and the parallelism degree of each layer gradually decreases. The required clock cycles are successively reduced in the tree structure search. To ensure stability and correctness, the total clock cycles of the tree structure search are set to 29 with a certain margin.

Simultaneously with the frame detection, the signal on the frequency points around $N/(2\times sps)$ and $N-N/(2\times sps)$ are used to estimate the initial SPO, which is calculated as 
\begin{equation}
\tau_{0}=\frac{sps}{2\pi} \arg \left[X\left(K\right) \cdot X^*\left(N-K\right) \right]
\end{equation}
where $K$ is equal to  $N/(2\times sps)$ . $\arg(\cdot)$ denotes calculating the angle of a complex value. $(\cdot)^*$ represents the conjugate value. $\textbf{X}$ is the frequency-domain signal after 144-point FFT and RRC. It should be noted that frame detection and SPO estimation should be performed simultaneously, which aligns the output times of frame detection and initial SPO estimation. Therefore, a delay of 13 clock cycles is added after the frame detection. When frame detection outputs the signal for the arrival of the burst frame, the judge allows the initial SPO to be fed into the feedback link of BM-FDTR.

The feedback loop of BM-FDTR is mainly composed of the FD interpolator (interp.), the Godard-based SPO estimation, the loop filter, and the numerically controlled oscillator (NCO). The Godard-based SPO estimation can be expressed as
\begin{equation}
e(n)= \sum_{k=(1-\alpha)K} ^{(1+\alpha)K-1} \operatorname{Im} [X(k) \cdot X^* (N-N/sps+k)] 
\label{eq:tao}
\end{equation}
where $\operatorname{Im}(\cdot)$ denotes the imaginary part of a complex value.$\alpha$ is the roll-off factor. Only $2\alpha N/sps$ frequency points are adopted to estimate the SPO. The summation in Eq. (\ref{eq:tao}) was implemented by a binary-tree structure, which takes 7 clock cycles. The $\tau$ estimated by Godard-based SPO estimation is fed into the loop filter to improve the stability and response of the feedback link, which can be calculated as
\begin{equation}
W(n) = k_p\cdot e(n) + k_i\cdot \sum_{l=0}^{n-1}e(n-l)
\label{loopfilter}
\end{equation}
where $k_p$ and $k_i$ are the coefficients of the proportional arm and integral arm, respectively. Then the control signal from the loop filter is sent to the NCO module, which produces the fractional and integer intervals for the interpolator. The oscillating signal $\eta(n)$ produced by the NCO can be calculated as
\begin{equation}
\eta(n) = \text{Mod}\left[\eta(n-1) - W(n),1\right]
\label{NCO}
\end{equation}
where $\text{Mod}(x,1)$ denotes the remainder of $x$ divided by $1$, which uses combinatorial logic rather than sequential logic to reduce the operation time. The fractional interval $\mu(n)$ is calculated as
\begin{equation}
\mu(n) = \eta(n-1)/W(n).
\label{frac_interval}
\end{equation}
Considering the complexity of division operation, timing violations, and resource usage, the default pipelining calculation clock cycles of the Vivado IP core are set to 39 clock cycles. The integer interval can be calculated as
\begin{equation}
m(n+1) = \left\{\begin{array}{ll}
m(n)-1, & \eta(n) - W(n) < -1\\
m(n), & -1 \leq \eta(n) - W(n) < 0\\
m(n) + 1, & \eta(n) - W(n) \geq 0 
\end{array}\right..
\label{int_interval}
\end{equation}
Finally, the calculated fractional and integer intervals are sent into the look-up table (LUT) to find the corresponding values of $e^{-2j\pi f \tau }$ for the FD interpolator, where $\tau = m(n) + \mu(n)$.

\begin{figure}[!t]
\centering
    {\includegraphics[width = \linewidth]{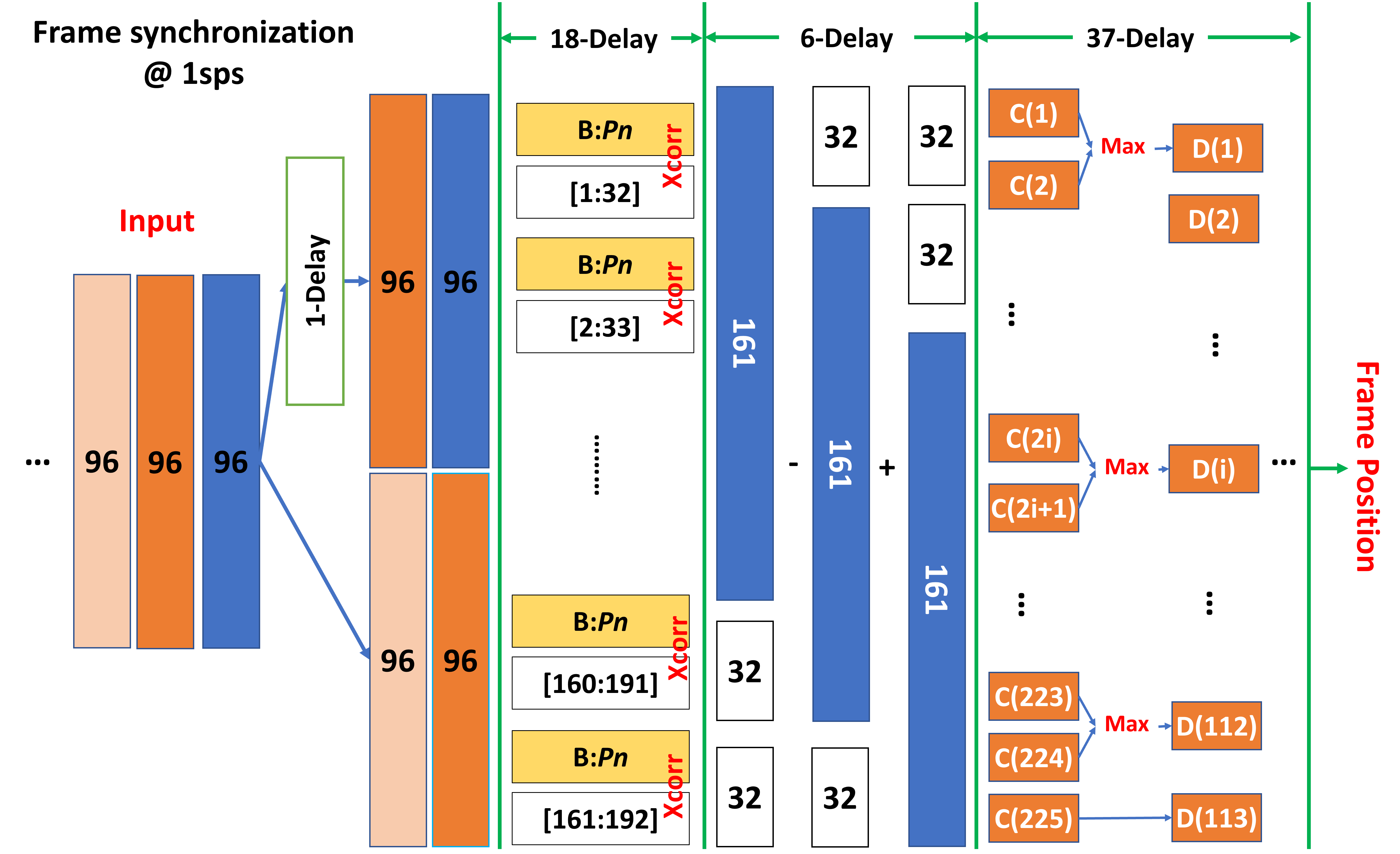}}
\caption{The real-time implementation of frame synchronization based on Preamble B.}
\label{fig:framesyn}
\end{figure}

\subsection{Real-Time Frame Synchronization}
Figure \ref{fig:framesyn} shows the real-time implementation of frame synchronization based on Preamble B. The input of frame synchronization is 1-sps signal, and its parallelism degree is 96. Two successive beats are reconstructed to one beat with a parallelism degree of 192. We can certainly find one received Preamble B in the reconstructed beats. $Pn$ in Preamble B is used for sliding cross-correlation (xcorr) with the reconstructed beat. $Pn$ is a sequence of $1$ and $-1$. Therefore, the xcorr can be implemented by addition and subtraction instead of multiplication to reduce DSP resources and timing issues. Moreover, the addition and subtraction for 32 samples can use the flow design to eliminate the possible wiring blockage problem. The 161 xcorrs can obtain 161 cross-correlation values, which take 18 clock cycles. 161 cross-correlation values are reconstructed to three matrixes with 225 values by adding two matrixes with 32 zeros. The three matrixes are multiplied by [1, 1, -1], and then are summed to generate one matrix with 225 values. This part takes 6 clock cycles.  Finally, the binary-tree search is employed to find the maximum value, which does the comparison between two values and reserves the bigger one for the next-stage comparison until the biggest value is found. The synchronization position is confirmed using the position of the biggest value, which takes 61 clock cycles. The synchronization position $p$ of the signal at 1.125 sps is obtained by $p = \lfloor p_1\times 1.125 \rfloor$ where $\lfloor \cdot \rfloor$ denotes the round-down operation. $p_1$ is the synchronization position of the signal at 1 sps \cite{wang2023non}. The fractional part of $p1 \times 1.125$ can be compensated by the FDTR.

\subsection{Real-Time BM-FDE}
\begin{figure}[!t]
\centering
    {\includegraphics[width = \linewidth]{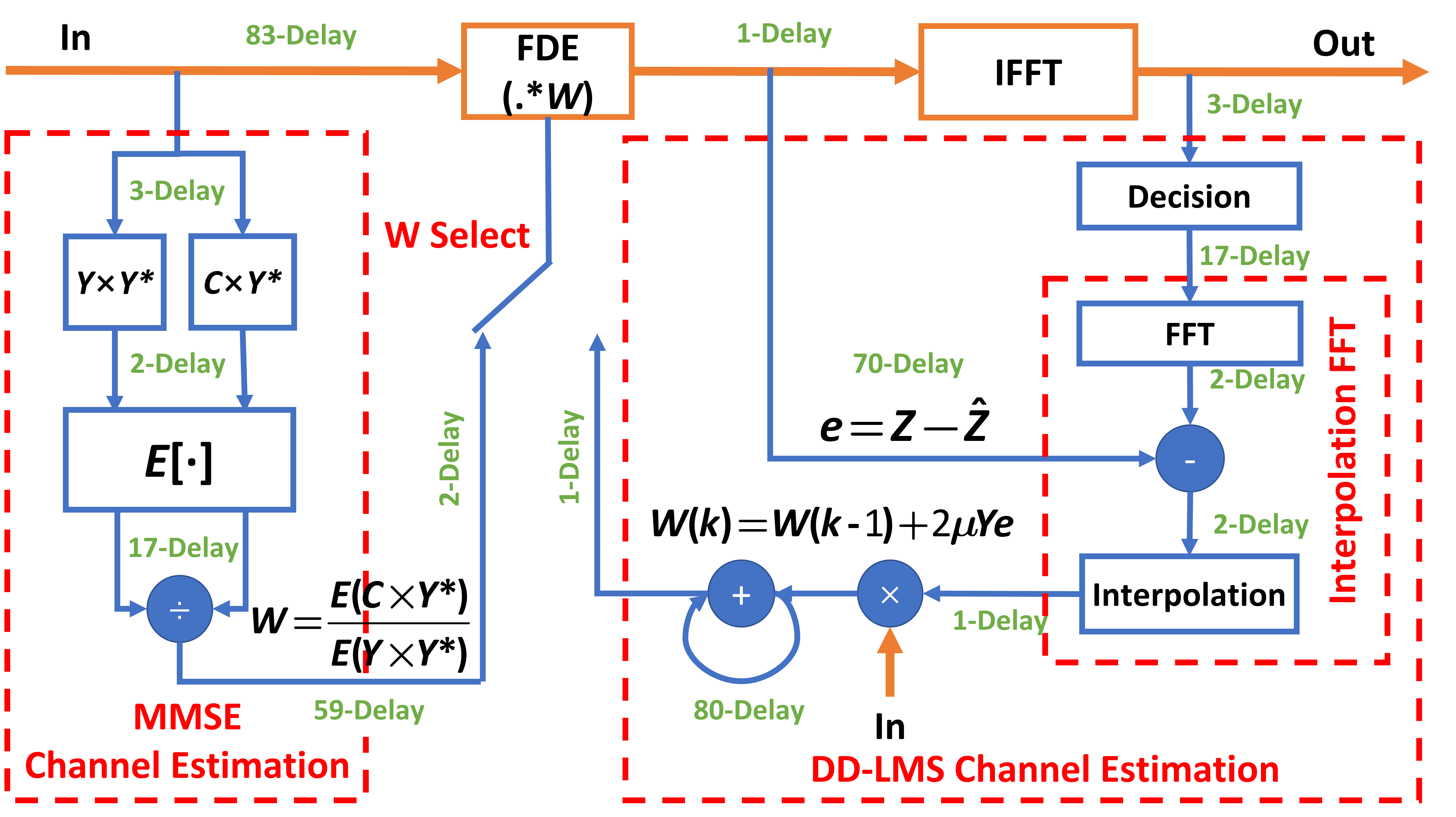}}
\caption{The real-time implementation of BM-FDE with feedforward channel estimation based on Preamble C.
}
\label{fig:fde}
\end{figure}

Figure \ref{fig:fde} shows the real-time implementation of BM-FDE with feedforward MMSE-based channel estimation based on Preamble C. The principle of BM-FDE has been demonstrated in the Ref. \cite{zhou2024burst}. In MMSE-based channel estimation, two mean operations $E(.)$ are performed on the $\mathbf{C}\times \mathbf{Y}^*$ and $\mathbf{Y}\times \mathbf{Y}^*$ where $\mathbf{Y}$ is 8-beat received signals after TR and synchronization. $\mathbf{C}$ is 8-beat transmitted Preamble-C signals. The mean operation can be simplified by adding the 8 multiplication outputs, and then shifting the addition results to the left by 3 bits, which saves the FPGA resources. The tap coefficients can be calculated by 
 \begin{equation}
\mathbf{W}_{\text{MMSE}} = E[\mathbf{C}\times \textbf{Y}^*]/E[\mathbf{Y}\times \textbf{Y}^*]
\label{eq:MMSE}
\end{equation}
where the division operation for two complex values consumes large resources. Therefore, the division operation requires 59 clock cycles to avoid timing problems and ensure enough FPGA resources. To compensate for channel response, the signals on the frequency points are multiplied by the corresponding tap coefficients. However, the channel response is dynamically changed. Therefore, the feedback DD-LMS algorithm can track and update the tap coefficients.

The DD-LMS algorithm has a high computational complexity due to the full-size FFT. This paper proposes a simplified DD-LMS algorithm based on the interpolation FFT. After FDE and IFFT, only 8 out of 128 time-domain symbols (i.e. one beat signal) are selected at equal intervals to implement the hard decision. 8-point FFT transfers the hard-decision symbols to the frequency-domain signal $\mathbf{\hat{Z}}$. To calculate the decision error, 8 signals $\mathbf{Z}$ out of 128 frequency points before the IFFT are selected at equal intervals, delaying 70 clock cycles. The decision error can be calculated by the $\mathbf{e}=\mathbf{Z}-\mathbf{\hat{Z}}$. The interpolation operation duplicates 16 times for each element of $\mathbf{e}$ to reconstruct 128 approximate decision errors. The computational complexity can be decreased at the expense of the accuracy. Then, the tap coefficients for the 128 frequency points can be updated by 
\begin{equation}
W_{n}(k) = W_{n}(k-1)+2\mu_{n}\times Y_n\times e_n
\label{eq:LMS}
\end{equation}
where $W_{n}$ is the tap coefficients for the $n$-th frequency point. $k$ denotes the $k$-th iteration. $\mu_{n}$ is the step size for the $n$-th frequency point. A delay of 80 clock cycles is required to align the $W_{n}(k)$ and $W_{n}(k-1)$. The MMSE-based Channel estimation is a feed-forward algorithm, which obtains the initial tap coefficients. Then, the tap coefficients can be updated by the feedback DD-LMS algorithm. The initial tap coefficients accelerate the convergence of the feedback DD-LMS algorithm.

\begin{figure}[!t]
\centering
    {\includegraphics[width = \linewidth]{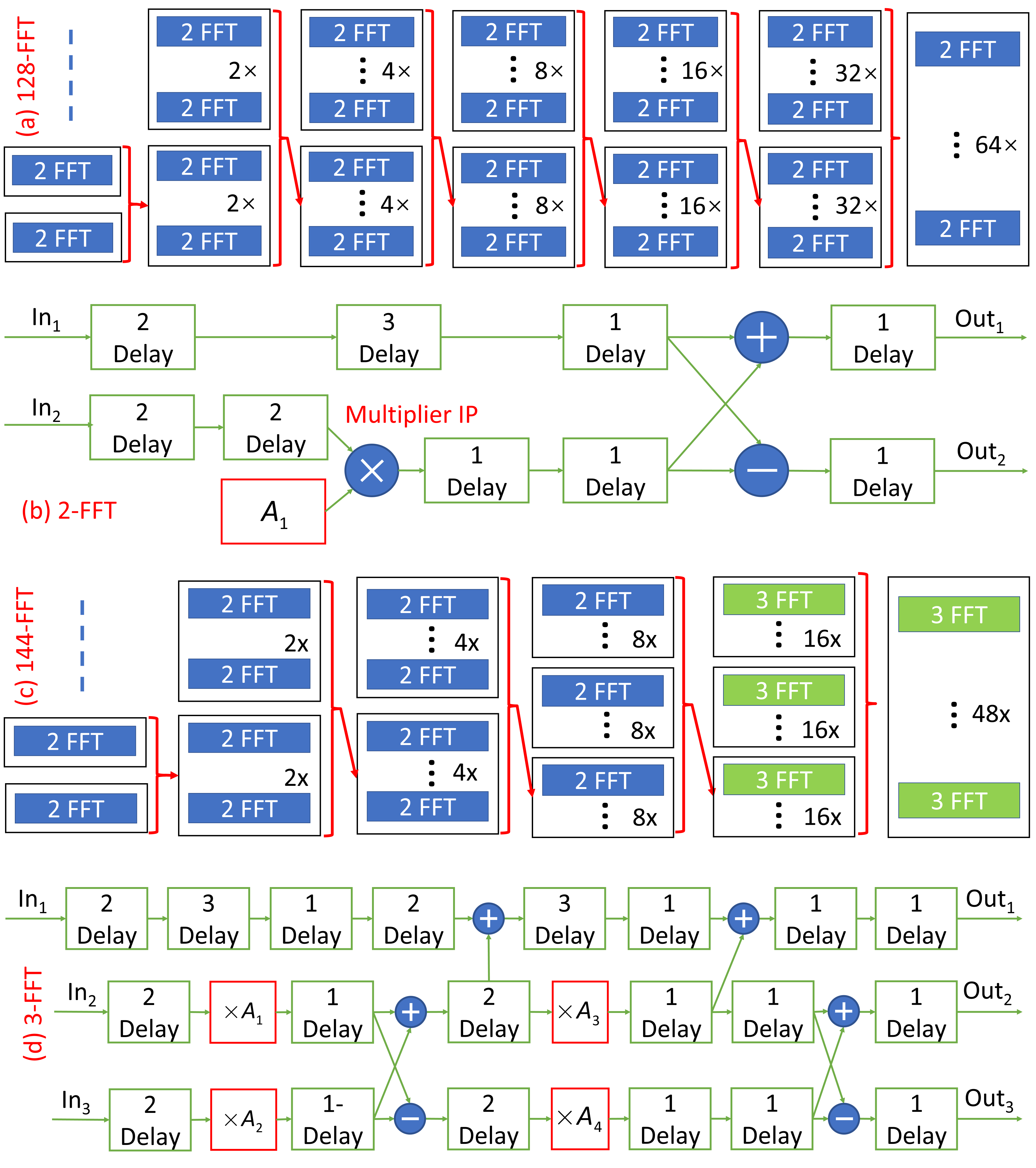}}
\caption{(a) Real-time implementation of 128-point FFT, (b) Real-time implementation of the radix-2 FFT, (c) Real-time implementation of 144-point FFT, (d) Real-time implementation of the radix-3 FFT.}
\label{fig:fft}
\end{figure}

\subsection{Real-Time 128/144-point FFT}
Figure \ref{fig:fft} (a) shows the real-time implementation of 128-point FFT based on the butterfly diagram. The 128-point FFT can be composed of 7 layers, and each layer has 64 radix-2 FFTs. Fig. \ref{fig:fft} (b) depicts the structure of radix-2 FFT. The coefficients $A_1$ for different radix-2 FFTs are set to $e^{j2\pi k/N}$ where $k$ are from 0 to $N/2-1$. $N$ is equal to $2^{n}$ where $n$ denotes the $n$-th radix-2 FFT layer. The $A_1$ for the radix-2 FFT in the first layer is 1, which only requires addition and subtraction with 4 clock cycles. For the other layers, the multiplication with a fixed coefficient of $A_1$ can employ the fixed-coefficient multiplier IP core, which uses the LUT resources to replace the DSP resources to avoid the excessive consumption of the DSP resources. Meanwhile, timing alignment for two paths in the radix-2 FFT is required in the pipeline. Two paths in the radix-2 FFT require 7 clock cycles, which can be further reduced when the FPGA resources are sufficient. The 128-point FFT requires 46 clock cycles. 

Figure \ref{fig:fft} (c) shows the real-time implementation of 144-point FFT. 144-point FFT consists of 4 radix-2 FFT layers and 2 radix-3 layers. Considering the module reusability, the four radix-2 FFT layers of 144-point FFT are similar to the first four radix-2 FFT layers of 128-point FFT. Then, two radix-3 FFT layers are implemented to complete the 144-point FFT. The real-time implementation of the radix-3 FFT is shown in Fig. \ref{fig:fft} (d). The radix-3 FFT consists of 3 computational paths. In the first path, delays of 14 clock cycles and 2 additions are implemented. In the second path, the coefficients $A_1$ for the first multiplier are set to $e^{j2\pi k/N}$ where $k$ are from 0 to $N/2-1$. $N$ is the number of points in one radix-3 FFT block. The coefficient $A_3$ for the second multiplier is set to $\text{cos}(2\pi/ 3)-1$. In the third path, the coefficients $A_2$ for the first multiplier are set to $e^{j4\pi k/N}$ where $k$ are from 0 to $N/2-1$. $N$ is the number of points in one radix-3 FFT block. The coefficient $A_4$ for the second multiplier is set to $j\text{sin}(2\pi/3)$. Significantly, there are delays of 3 clock cycles for each multiplier to ensure the same delays for the three paths. In the second and third paths, delays of 14 clock cycles, 2 fixed-coefficient multipliers, and 2 additions are implemented. In conclusion, one 144-point FFT requires 53 clock cycles. For both 128-point and 144-point FFTs, the outputs are Hermitian conjugation when the inputs are real values. Therefore, the resources for real-valued FFT can be further reduced \cite{sorensen1987real}.

\begin{table*}[bp]
 \centering
\caption{ \centering The FPGA resources for the main algorithms of BM-DSP before and after optimization. The dataY/ dataX denotes that dataX is the data before optimization and dataY is the data after optimization.}
 \label{tab:resource}
 \renewcommand{\arraystretch}{1.5}
\scalebox{1}{ 
  \centering
    \begin{tabular}{ccccccc} \hline 
         Type/Module&  FFT&  RRC&  TR& FDE&  Frame Sync.&  Ratio to Total(\%)\\ \hline 
         LUT&  140,085/229,591& 3,207& 13,028& 282,940/367,093&  131,497&  51.55/62.97\\ 
         LUTRAM&  45,230/46,582& 0& 3,775/3,762& 11,748/65,853&  7,775/9,163&  12.54/22.25\\ 
         FF&  155,679/285,729&  9,492&   16,466&  564,356/694,298&  294,797&  46.29/57.71\\ 
 BRAM& 0& 0&  75.5&0& 9&12.52\\ 
 DSP& 2,137/2,573& 292&  473& 1,170/2,724& 1& 59.55/88.12\\ \hline
    \end{tabular}
} 
\end{table*}

\section{Experimental setup and results}\label{Section3}
\begin{figure}[!t]
\centering
    {\includegraphics[width = \linewidth]{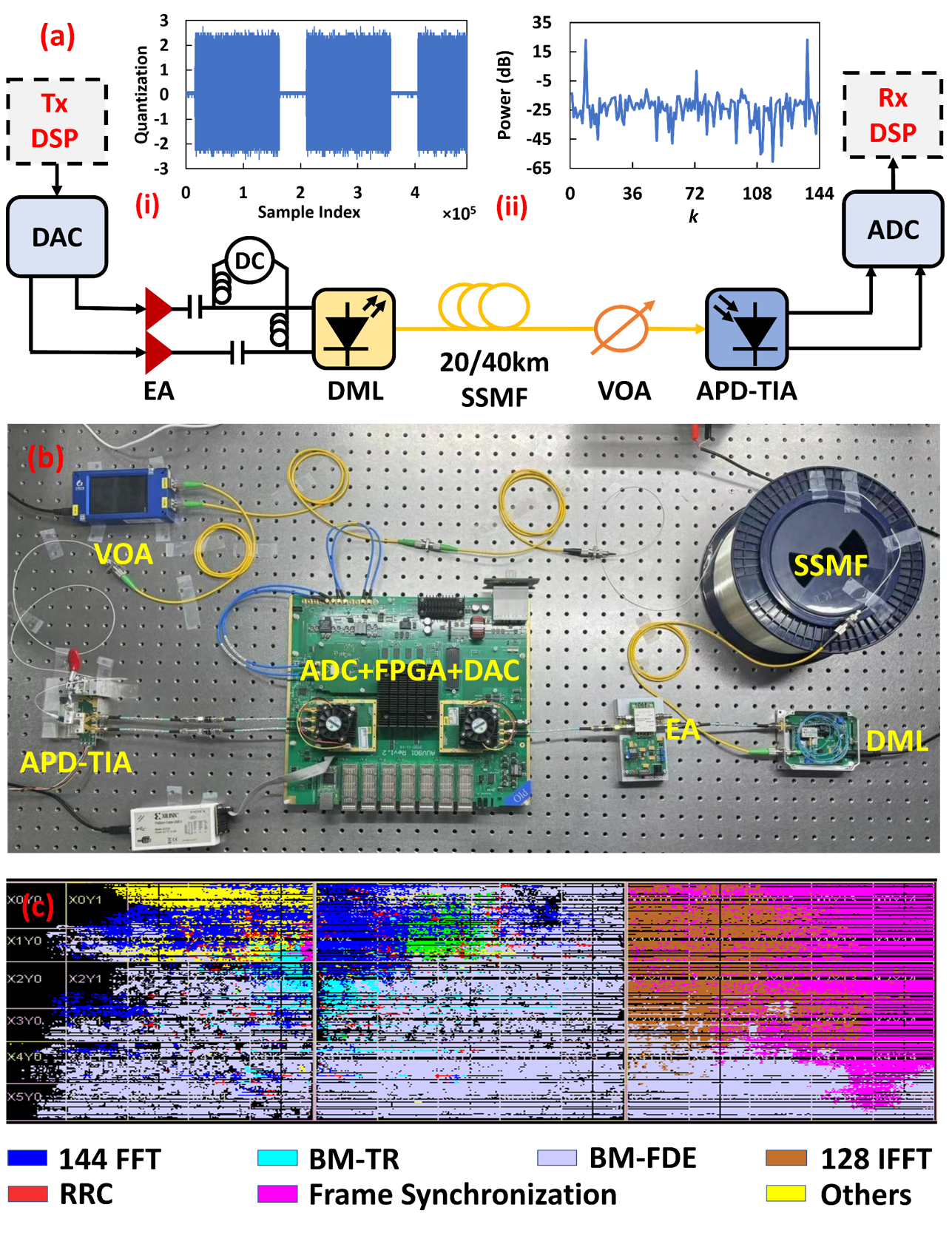}}
\caption{(a) The block diagram of experimental setups for the 25Gbit/s OOK using 10G-class DML and APD-TIA. Inset (i) shows the transmitted electrical signal. Inset (ii) depicts the spectrum of the received tone signals. (b) The real-time experimental platform. (c) The FPGA resource distribution.}
\label{fig:system}
\end{figure}

Figure \ref{fig:system} (a) shows the block diagram of the experimental setups for 25Gbit/s OOK using 10G-class directly modulated laser (DML) and avalanche photodiode with trans-impedance amplifier (APD-TIA). The DSP at the transmitter generated the digital OOK signal as shown in Fig. \ref{fig:dspflow}. Before the first-in-first-out (FIFO) module, the parallelism degree is 108, and the clock frequency is set to 261.12MHz. After the FIFO module, the parallelism degree is 128, and the clock frequency is changed to 220.32MHz. Therefore, the sampling rate of DAC is 28.2GS/s, which is equal to the parallelism degree multiplied by the clock frequency. After the DAC, the analog signal was generated, which is shown in the inset (i) of Fig. \ref{fig:system} (a). Two frames with a big gap were used to simulate the burst scenario. In the real scenario, the data length is longer, and the gap is smaller, which must adhere to the ITU-T standards. An electrical amplifier amplified the analog signal.  Then, the amplified analog signal was modulated on a 1328nm optical carrier by a 10G-class DML. 

The generated optical signal was launched into 20km and 40km standard single-mode fiber (SSMF) with the launch optical power of $\sim$4.5dBm. A variable optical attenuator (VOA) was used at the receiver to adjust the received optical power (ROP). A 10G-class receiver optical subassembly (ROSA) based on the APD-TIA was used to convert the optical signal to an electrical signal. The electrical signal was sent to an analog-to-digital converter (ADC) with a sampling rate of 28.2GS/s to convert the analog signal to a digital signal. Finally, the real-time BM-DSP based on FPGA processed the digital signal, as shown in Fig. \ref{fig:dspflow}. The inset (ii) of Fig. \ref{fig:system} shows the spectrum of the received tone signals acquired by the integrated logic analyzer (ILA), which reveals that there are two peaks at two frequency points. Fig. \ref{fig:system} (b) shows the photo of the real-time experimental platform. 

After synthesis, the distribution of the FPGA resources is shown in Fig. \ref{fig:system} (c). The frame synchronization and BM-FDE use
a large of FPGA resources. The purple part shows the FPGA resources used for frame synchronization, which requires complex sliding cross-correlation. The gray part represents the resources used for BM-FDE, which requires complex MMSE initializing and the DD-LMS updating algorithms. Table \ref{tab:resource} shows the FPGA resources for the main algorithms of BM-DSP before and after optimization. After optimization, the consumption ratio of LUT is 51.55\%. The consumption ratio of random access memory for LUT (LUTRAM) is 12.54\%. The consumption ratio of FF is 46.29\%. The consumption ratio of block RAM (BRAM) is 12.52\%. The consumption ratio of DSP is 59.55\%. Owing to the improved implementations for FFT, MMSE, and DD-LMS algorithms decrease the DSP resources by 28.57\%, enabling the loading of real-time BM-DSP in the FPGA with the limited DSP resources.

\begin{figure}[!t]
\centering
    {\includegraphics[width = 0.9\linewidth]{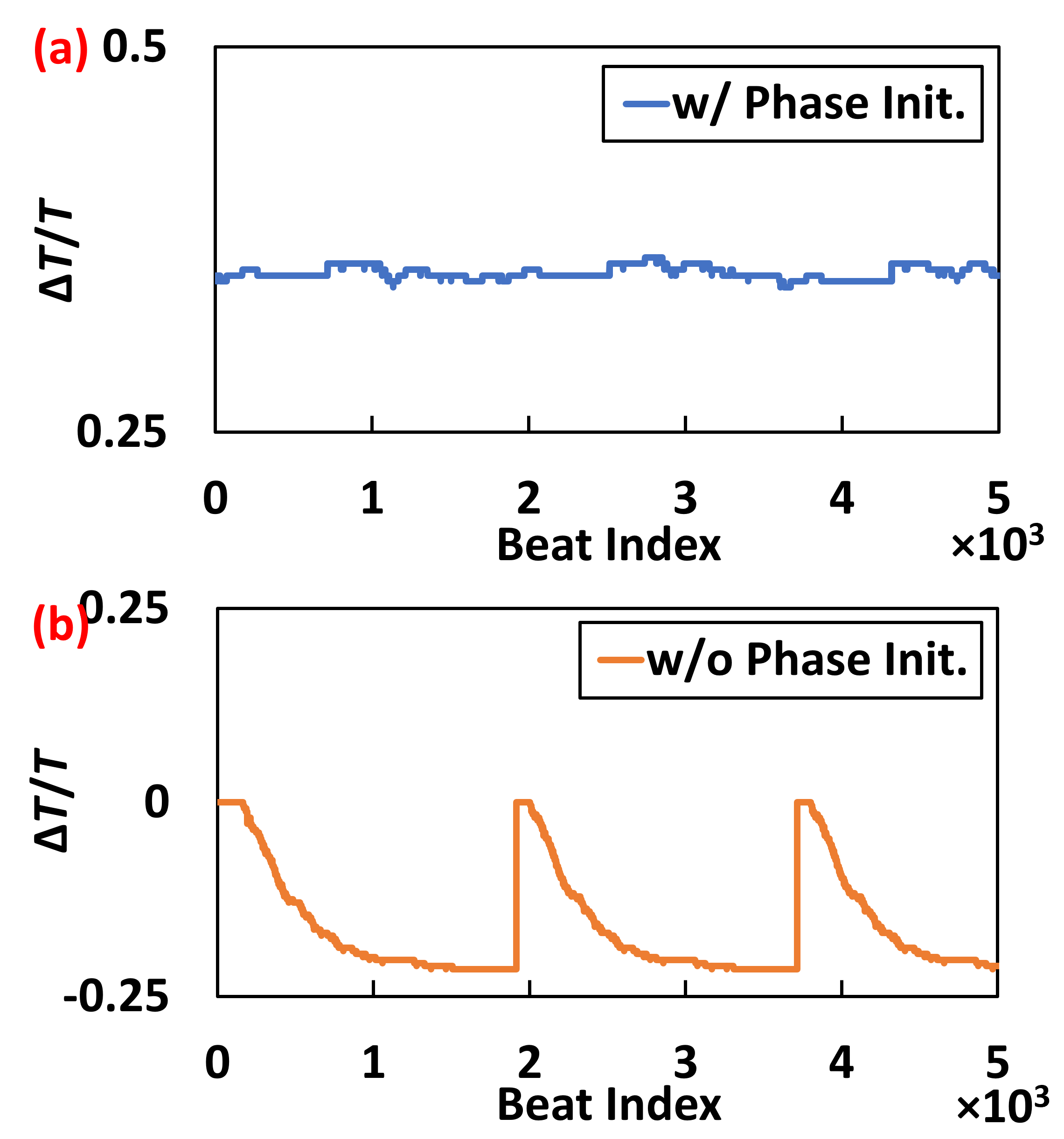}}
\caption{(a) The SPO estimated by the BM-FDTR with the SPO initialization. (b) The SPO estimated by the FDTR without the SPO
initialization.}
\label{fig:trresult}
\end{figure}

\begin{figure}[!t]
\centering
    {\includegraphics[width = 0.9\linewidth ]{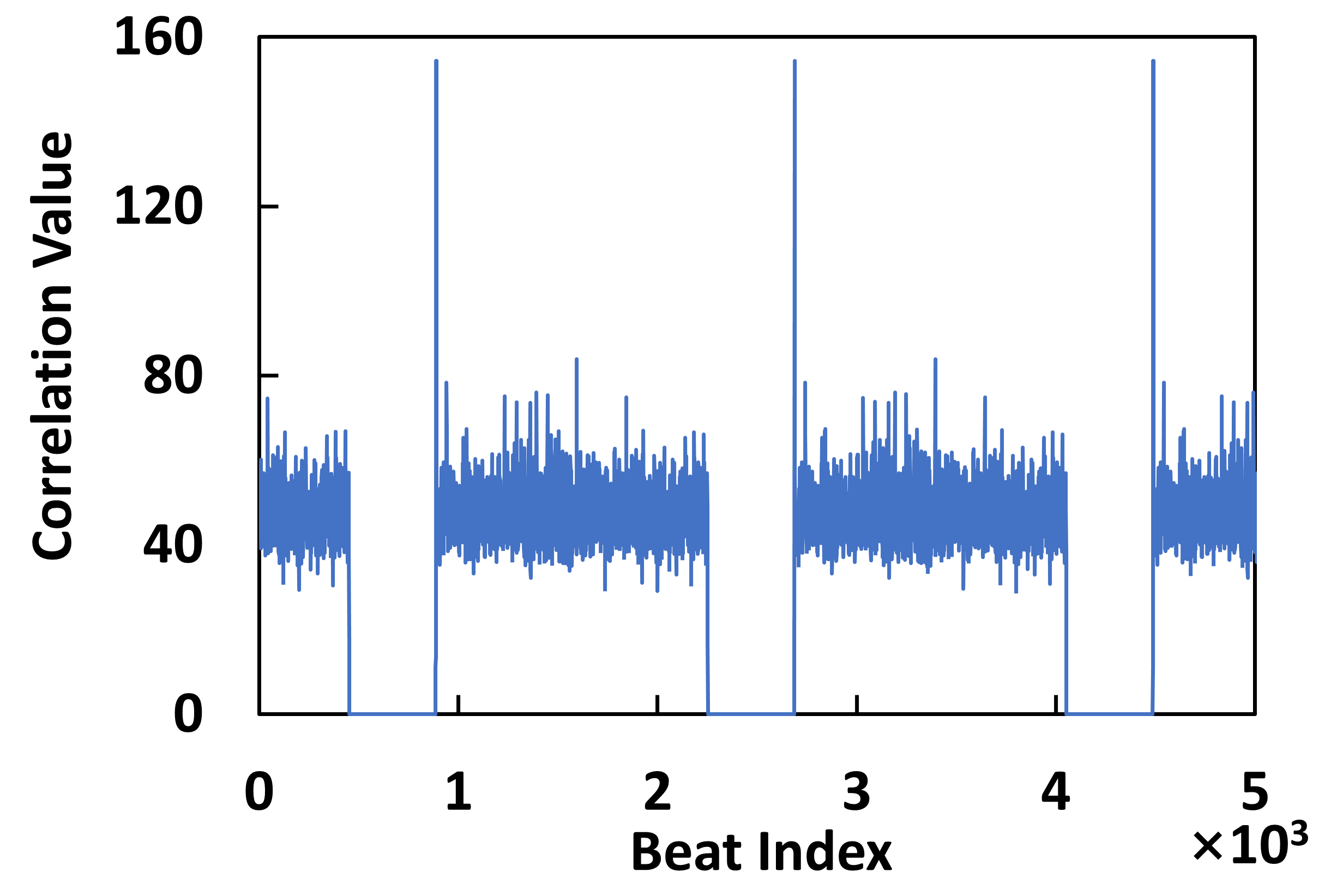}}
\caption{The correlation values between the received signal and Preamble B versus beat index for the frame synchronization.}
\label{fig:framresult}
\end{figure}

An ILA was used to collect parameters such as SPO for BM-FDTR, autocorrelation values for frame synchronization, and MSE for BM-FDE to analyze the statuses of the mentioned real-time algorithms. Fig. \ref{fig:trresult} (a) shows the estimated SPO by the BM-FDTR with the SPO initialization. Fig. \ref{fig:trresult} (b) depicts the estimated SPO by the FDTR without the SPO initialization. The feed-forward algorithm based on Preamble A estimated the initial SPO. When an accurate SPO initializes the BM-FDTR, the BM-FDTR can track the SPO rapidly. The FDTR without SPO initialization required more signal beats to track the SPO. In general, the signal beats within tracking time have worse performance than those within converged time. Therefore, the BM-FDTR with SPO initialization performs better than without SPO initialization within tracking time. The exact Preamble C should be used to estimate the tap coefficients of BM-FDE. After the BM-FDTR, the frame synchronization should be implemented to confirm the beginning of the Preamble C. Fig. \ref{fig:framresult} shows the correlation values between the received signal and the Preamble B versus beat index for the frame synchronization. The correlation values have one peak value in one frame, which is much higher than the other values. When the peak of correlation values is detected, an accurate frame position is confirmed owing to the high peak-to-noise power ratio. Therefore, the frame synchronization based on Preamble B can obtain the accurate frame position to confirm the beginning of Preamble C.

\begin{figure}[!t]
\centering
    {\includegraphics[width = 0.9\linewidth]{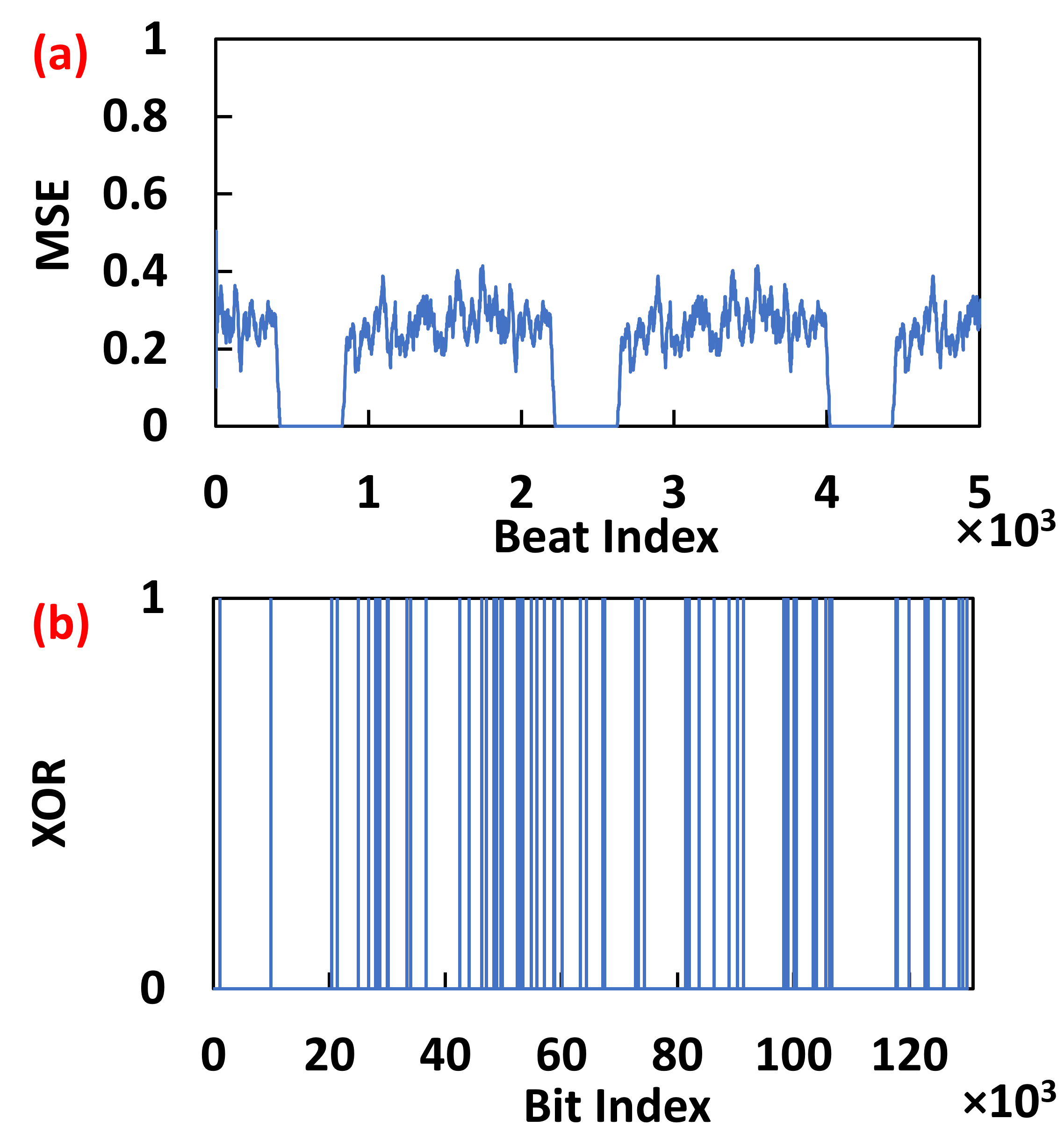}}
    \caption{(a) The MSE of the BM-FDE when the feed-forward MMSE algorithm initializes and the feedback DD-LMS algorithm updates the tap coefficients. (b) Distribution of error bits for one burst frame at the ROP of $-23$dBm. }
\label{fig:fderesult}
\end{figure}

Figure \ref{fig:fderesult} (a) shows the MSE of the BM-FDE when the feed-forward MMSE algorithm initializes the tap coefficients based on Preamble C and the feedback DD-LMS algorithm updates the tap coefficients. The feed-forward MMSE algorithm based on Preamble C accurately estimates tap coefficients. The MSE of the DD-LMS algorithm is stable when the tap coefficients are initialized. Therefore, the tap coefficients of the BM-FDE are quickly converged when it is initialized by the estimated tap coefficients. Fig. \ref{fig:fderesult} (b) depicts the distribution of the error bits for one burst frame at the ROP of $-23$dBm. The error bits' distribution is uniform, also indicating that the tap coefficients estimated by the feed-forward MMSE algorithm are exact. When the tap coefficients are not initialized, there are more error bits at the beginning of the burst frame. When the feed-forward MMSE algorithm initializes the tap coefficients, the BM-FDE performs excellently on processing the burst signal.

\begin{figure}[!t]
\centering
    {\includegraphics[width = 0.9\linewidth]{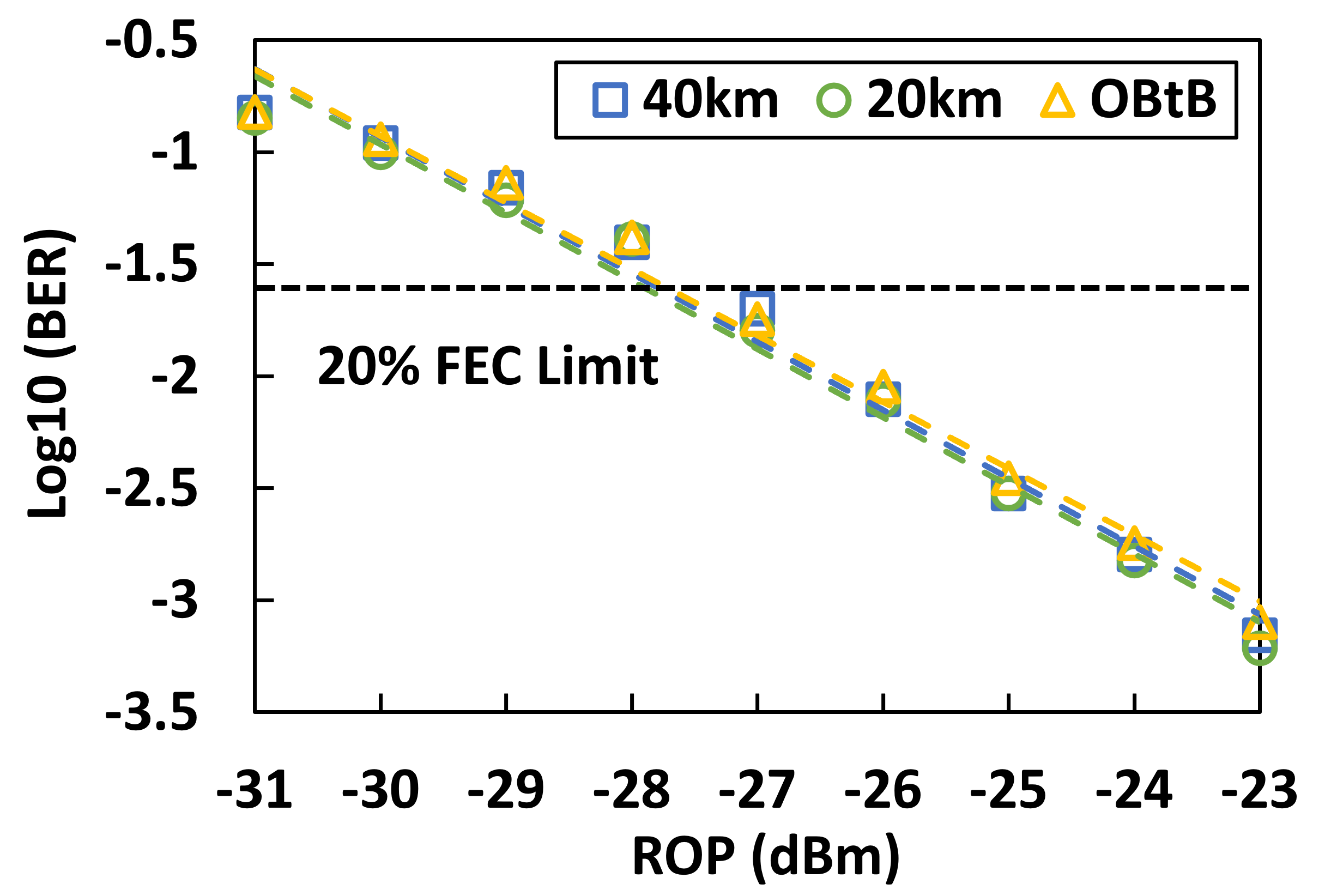}}
\caption{ The BER versus ROP for 25Gbit/s OOK burst reception after OBtB, 20km, and 40km SSMF transmission. The dashed line denotes the 20\% FEC limit.}
\label{fig:ber}
\end{figure}

Figure \ref{fig:ber} shows the BER versus ROP for 25Gbit/s OOK burst reception after OBtB, 20km, and 40km SSMF transmission. The dashed line denotes the $20\%$ FEC limit. There is almost the same BER performance for 25Gbit/s OOK burst reception after OBtB, 20km, and 40km SSMF transmission, which can achieve the $20\%$ FEC limit at the ROP of $\sim-27.5$dBm. Therefore, the slight chromatic dispersion of the O band optical signal barely influences the BER performance of 25Gbit/s OOK signal after the BM-DSP. When the launch optical power is set to $\sim 4.5$dBm, the optical power budget is approximately 32dB, which can satisfy the maximum optical path loss of Class C+ for 50G PON. In conclusion, the proposed BM-DSP can effectively process the 25Gbit/s OOK upstream signal.

\section{CONCLUSION} \label{Section4}
In this paper, we implement real-time BM-DSP for burst reception of 25Gbit/s OOK, which can meet the asymmetric-mode 50G PON demand. The real-time BM-DSP including the BM-FDTR and BM-FDE can be fast converged based on the 42ns designed preamble. Meanwhile, the improved implementations for FFT, MMSE, and DD-LMS algorithms decrease the DSP resources by 28.57$\%$, enabling the loading of real-time BM-DSP in the FPGA with the limited DSP resources. After up to 40km SSMF transmission, the optical power budget was approximately 32dB when the BM-DSP was employed, which can satisfy the maximum optical path loss of Class C+. The reason why we realize 25Gbit/s OOK burst reception is the sampling rate limitation of the DAC and ADC. However, the algorithm implementation can be also applied in 50Gbit/s OOK burst reception in theory. In conclusion, the real-time implementation of BM-DSP can guide the DSP ASIC design for 50G PON.

\section*{Funding} This work was supported in part by the National Key R\&D Program of China under Grant 2023YFB2905700, in part by the National Natural Science Foundation of China under Grant 62371207 and Grant 62005102, in part by the Young Elite Scientists Sponsorship Program by CAST under Grant 2023QNRC001, and in part by the Hong Kong Research Grants Council GRF under Grant 15231923.

\bibliography{sample}

\end{document}